\documentclass[a4paper,12pt]{article}
\usepackage{epsfig}
\usepackage{epstopdf}
\usepackage[sumlimits,intlimits,namelimits]{amsmath} 
\usepackage{dsfont}

\usepackage[english]{babel}

\usepackage{graphicx}
\usepackage{xcolor}

\usepackage{geometry}
\usepackage{braket}

\usepackage[square,sort,comma,numbers,compress]{natbib}
\definecolor{darkgreen}{rgb}{0,0.8,0}
\usepackage[pdftex,bookmarks=true,linktocpage=true,colorlinks=true,
    linkcolor=red,          
    citecolor=darkgreen,        
    filecolor=magenta,      
    urlcolor=cyan           
    ]{hyperref}

\usepackage{tabu}

\allowdisplaybreaks[1]

\makeatletter
\def\fmslash{\@ifnextchar[{\fmsl@sh}{\fmsl@sh[0mu]}}
\def\fmsl@sh[#1]#2{%
  \mathchoice
    {\@fmsl@sh\displaystyle{#1}{#2}}%
    {\@fmsl@sh\textstyle{#1}{#2}}%
    {\@fmsl@sh\scriptstyle{#1}{#2}}%
    {\@fmsl@sh\scriptscriptstyle{#1}{#2}}}
\def\@fmsl@sh#1#2#3{\m@th\ooalign{$\hfil#1\mkern#2/\hfil$\crcr$#1#3$}}
\makeatother
\DeclareMathOperator{\Tr}{Tr}

\numberwithin{equation}{section}
\begin{document}
\begin{titlepage}
\begin{flushright}
SI-HEP-2014-05 \\[0.2cm]
\end{flushright}

\vspace{1.2cm}
\begin{center}
{\Large\bf 
Improved Estimates for the Parameters \\[2mm] of the Heavy Quark Expansion}
\end{center}

\vspace{0.5cm}
\begin{center}
{\sc Johannes Heinonen} and {\sc Thomas Mannel } \\[0.1cm]
{\sf Theoretische Elementarteilchenphysik, Naturwiss.- techn. Fakult\"at, \\
Universit\"at Siegen, 57068 Siegen, Germany}
\end{center}

\vspace{0.8cm}
\begin{abstract}
\vspace{0.2cm}\noindent
We give improved estimates for the non-perturbative parameters appearing in the heavy quark expansion 
for inclusive decays. While the parameters appearing in low orders of this expansion can be extracted from data, 
the number of parameters in higher orders proliferates strongly, making a determination of these parameters 
from data impossible. Thus, one has to rely on theoretical estimates which may be obtained from an insertion of intermediate states. 
In this paper we refine this method and attempt to  estimate  the uncertainties of this approach.
\end{abstract}

\end{titlepage}

\newpage
\pagenumbering{arabic}
\section{Introduction}
The heavy mass expansion has become a mature method for the calculation of inclusive decay rates of 
heavy hadrons. It relies on the fact that inclusive decay rates and spectra for the decays of $B$ hadrons 
can be computed in a series in powers of $\Lambda_{\rm QCD} / m_b$, where the underlying technology 
is the operator product expansion (OPE) in QCD \cite{Shifman:1984wx,Chay:1990da,Bigi:1992su,Bigi:1993fe}. 

The non-perturbative input at each order is determined by forward matrix elements of local operators, which 
themselves have again a heavy mass expansion. Up to and including terms of the order
$(\Lambda_{\rm QCD} / m_b)^3$, there appear in total four non-perturbative quantities, which are the 
kinetic energy parameter $\mu_\pi$,  the chromo-magnetic parameter $\mu_G$, the Darwin term
$\rho_D$  and the spin-orbit term $\rho_{LS}$.  
   
The data and the theoretical framework for inclusive semileptonic decay rates and the lepton energy 
and hadronic mass spectra have developed to a point that these non-perturbative parameters can be determined 
or at least strongly constrained. Overall this has lead to a precision determination of $V_{cb}$ with a 
relative uncertainty of less than 2\% \cite{Benson:2003kp,Gambino:2013rza}. 

However, going beyond the order $(\Lambda_{\rm QCD} / m_b)^3$ requires many more non-perturbative 
parameters. In fact, the number of independent parameters proliferates significantly; at order 
$(\Lambda_{\rm QCD} / m_b)^4$ we have already nine parameters \cite{Dassinger:2006md}, 
while at $(\Lambda_{\rm QCD} / m_b)^5$
one finds eighteen \cite{Mannel:2010wj}. There is a factorial growth of this number, which has lead to speculations that the 
heavy quark expansion is asymptotic, just as the usual perturbative expansion~\cite{Shifman:1994yf,Shifman:2000jv}.  

In order to get an estimate for the effects of the higher orders in the heavy mass expansion it is thus important 
to get a reliable estimate of these higher-order contributions. To this end, one needs to estimate the forward 
matrix elements of higher-dimensional operators with heavy quarks. 

In previous papers ideas have been developed how to get an estimate for such matrix 
elements \cite{Mannel:2010wj}. 
The methods employed are based on a product of two operators, which on the one hand can be 
evaluated by inserting a set of intermediate states, while on the other hand one may perform an OPE. 
Truncating the infinite sum over intermediate states after the lowest-lying state 
(``lowest-lying  state saturation ansatz'', LLSA)  one obtains the higher order matrix elements in terms 
of $\mu_\pi$, $\mu_G$, $\rho_D$ and $\rho_{LS}$. 

In the present paper we  elaborate on this ideas further. First of all, we develop a systematic way to define the 
LLSA.
%
%
We show that up to order $(\Lambda_{\rm QCD} / m_b)^5$ the non-perturbative matrix elements can be expressed in LLSA by the
four parameters mentioned above.
Secondly, we discuss the uncertainties induced 
by the LLSA; this cannot  be done rigorously, but on the basis of a simple model one can get an idea on the 
precision of LLSA. 

In the next section we use the OPE to derive formulae which allow to calculate the LLSA. The arguments go 
very much along the lines of  \cite{Mannel:2010wj}, where a similar approach has been considered, however, 
not in a systematic way. We will also comment on some features of this approach. In section~\ref{sec:GSS} we show how to use the formalism developed in section~\ref{sec:framework} by deriving estimates for the higher-order matrix elements up to order $(\Lambda_{\rm QCD} / m_b)^5$ in terms of only four 
independent parameters and by giving numerical estimates for these. Finally, in section~\ref{sec:uncertainties} we discus the uncertainties of
the LLSA truncation by setting up a simple model to estimate the 
systematical uncertainties of LLSA. Then we will conclude and summarize our results in section~\ref{sec:conclusion}.

\section{Framework \label{sec:framework}} 
We are interested in deriving an expression for an expectation value $\braket{B|O|B}$ in terms of matrix elements of lower-dimensional operators. The analog in non-relativistic quantum mechanics for an operator of the form $O = O_1 O_2$ is given by using the completeness of states,
\begin{equation}
\braket{\psi| O | \psi} = \sum_n \braket{\psi| O_1|n}\braket{n|O_2 |\psi}.
\end{equation}
However, in a quantum field theory the operators can be local operators $O_1 (x)$ and $O_2 (x)$, where in general a product taken at 
the same space-time point $O_1 (x) O_2(x)$ is ill defined and needs renormalization.  
We will consider this problem very similarly to what was worked out in~\cite{Mannel:2010wj}, where it was treated less systematically. 
To set up our framework we  first distinguish between spatial and time derivatives. The spatial 
derivative is defined as 
\begin{equation}
D_\mu^\perp = g_{\mu \nu}^\perp   D^\nu \qquad \text{ with } \qquad  g_{\mu \nu}^\perp = g_{\mu \nu} - v_\mu v_\nu  
\end{equation}
while the time derivative is just  given by $ D_t := v\cdot D$, where $v^\mu$ denotes the four-velocity of the $B$-meson.
We consider first operators with a chain of only spatial derivatives and define 
\begin{subequations}
\begin{eqnarray}
\mathcal P_1 &=& (iD_{\mu_1}^\perp) (iD_{\mu_2}^\perp) ...  (iD_{\mu_m}^\perp),  \\
\mathcal P_2 &=&  (iD_{\nu_1}^\perp) (iD_{\nu_2}^\perp) ...  (iD_{\nu_n}^\perp).
\end{eqnarray}
\end{subequations}
The Lorentz indices in these equations might be contracted or left open in what follows.

It is useful to introduce a fictitious heavy quark $Q$ with a mass much larger than the $b$ quark mass
, $m_Q \gg m_b$.
Using this we may form the operators
\begin{subequations}
\begin{eqnarray}
O_1 (x) &=& \bar{b}(x) \mathcal P_1  Q (x),  \\
O_2 (y) &=& \bar{Q}(y)   \mathcal P_2 {\Gamma} b (y), 
\end{eqnarray} 
\end{subequations}
where $\Gamma$ is an arbitrary Dirac matrix, which can be chosen to only appear in the second operator.
As discussed in~\cite{Mannel:2010wj}, one considers the Fourier transform of the forward matrix element of the time-ordered product 
(without loss of generality set $y=0$) 
\begin{equation}
{T}(v\cdot q) := \int d^4 x \, e^{ i \, (v\cdot x)  \, (v \cdot q)}
\braket{ B(p_B) | \mathcal T \big\{ O_1 (x) O_2 (0) \big\} | B(p_B) }
\end{equation}  
and performs the standard steps for a heavy mass expansion: We redefine the quark fields as 
\begin{equation}
b(x) = e^{-i m_b v \cdot  x} b_v (x) 
\qquad \text{ and } \qquad 
 Q(x) = e^{-i m_Q v \cdot x} Q_v (x), 
\end{equation} 
which suggests to define the parameter $\omega = (v\cdot q) + m_b - m_Q$ and thus
\begin{equation}  \label{eq:Tomega}
T (\omega) = \int d^4 x \, e^{ i \,  \,  (v\cdot x) \omega}
\Braket{ B(p_B) | \mathcal T \big\{ \bar{b}_v (x)  \mathcal P_1 Q_v(x)   \,  
     \bar{Q}_v (0)  \mathcal P_2  {\Gamma} b_v(0)  \big\} | B(p_B) }.
\end{equation} 
Now we insert a complete set of intermediate states and use that momentum is the generator of translations, such that for any operator $O(x)=e^{i P\cdot x} O(0) e^{-i P\cdot x }$. 
In the rest frame of the decaying $B$ meson one finds\footnote{Here and in what follows, a field without a space-time argument is to be taken at $x=0$.}  
\begin{eqnarray} \label{eq:intState}
T (\omega) &=& \sum_n \frac{i (2 \pi)^3 \delta^3 (p_n^\perp)}{\omega - \epsilon_n + i \varepsilon} 
\Braket{ B(p_B) | \bar{b}_v    \mathcal P_1{} Q_v | n } \, \Braket{ n |  \bar{Q}_v    \mathcal P_2 {\Gamma}  b_v  | B(p_B) }
\\ \nonumber   
&+& \sum_n \frac{i (2 \pi)^3 \delta^3 (p_n^\perp)}{\omega + 2 (M_Q - M_B) + \epsilon_n + i \varepsilon} 
\Braket{ B(p_B) |  \bar{Q}_v  \mathcal P_2 {\Gamma}  b_v    | n } \, \Braket{ n |  \bar{b}_v   \mathcal P_1 {} Q_v | B(p_B) }.
\end{eqnarray}  	 
Here $\epsilon_n$ are the excitation energies, defined by the masses $M_n$ of the excited 
$Q$ hadron states as  $M_n = M_Q + \epsilon_n$, where $M_Q$ is the mass of the pseudoscalar ground state 
$Q$ meson.  
The second term is a contribution with intermediate $B$ and $Q$ states; in the limit of infinite quark mass $m_Q$ 
this contribution vanishes.  

For sufficiently large $\omega$ (i.e. $|\omega| \gg \Lambda_{\rm QCD}$) we can perform an OPE for the correlator 
$T(\omega)$ in \eqref{eq:Tomega}. The tree-level term of this OPE is simply obtained form contracting the intermediate $Q$ propagator. 
We are interested in the  limit $m_Q \to \infty$, in which case we may replace the propagator by the 
static propagator in the external gluon field within the $B$ meson
\begin{equation} \label{eq:OPE}
T (\omega) = \Braket{ B(p_B) | \bar{b}_v   \mathcal P_1  \left( \frac{i}{\omega+iv\cdot D  +i\varepsilon }
\right) \left(
   \frac{1+\fmslash{v}}{2} \right) \mathcal P_2  \Gamma       b_v    | B(p_B) }. 
\end{equation}   
Combining eq.~\eqref{eq:intState} and~\eqref{eq:OPE}  we obtain the relation 
\begin{eqnarray}  \label{master} 
&& 
\sum_n \frac{i (2 \pi)^3 \delta^3 (p_n^\perp)}{\omega - \epsilon_n + i \varepsilon} 
\braket{ B(p_B) | \bar{b}_v   \mathcal P_1 {}Q_v | n } \, \Braket{ n |  \bar{Q}_v   \mathcal P_2  {\Gamma} b_v   | B(p_B) }
\\ \nonumber && \qquad \qquad 
= \Braket{ B(p_B) | \bar{b}_v   \mathcal P_1  \left( \frac{i}{\omega+ iv\cdot D +i\varepsilon
}  \right) \left(  \frac{1+\fmslash{v}}{2} \right) \mathcal P_2 \Gamma    
    b_v   | B(p_B) }
\end{eqnarray} 
in the $m_Q \to \infty$ limit. For large $\omega$ this formula can be expanded in inverse powers of $\omega$ 
yielding the final relation
\begin{eqnarray} \label{eq:master1}
&& 
\sum_{k = 0}^\infty \sum_n   (2 \pi)^3 \delta^3 (p_n^\perp) \left( \frac{-\epsilon_n}{\omega} \right)^k 
\Braket{ B(p_B) | \bar{b}_v   \mathcal P_1{} Q_v | n } \, \Braket{ n |  \bar{Q}_v  \mathcal P_2 {\Gamma}  b_v  | B(p_B) }
\\ \nonumber && \qquad \qquad 
= \sum_{k = 0}^\infty  \Braket{ B(p_B) | \bar{b}_v  \mathcal P_1  \left(\frac{iv\cdot D}{\omega}  \right)^k  
\left( \frac{1+\fmslash{v}}{2}  \right)\mathcal P_2 \Gamma    
    b_v   | B(p_B) }.
\end{eqnarray} 
This equations establishes our goal of relating a matrix element of the schematic form $\braket{B| \mathcal P_1 \mathcal P_2 |B}$ to (a sum of) products of matrix elements of lower-dimensional operators $\braket{B|\mathcal P_1|n}\braket{n|\mathcal P_2|B}$.  In the following sections we will demonstrate how to put this equation to use and obtain estimates for matrix elements up to order $(\Lambda_{\rm QCD} / m_b)^5$.

Before we go on to show how the master equation~\eqref{eq:master1} is applied for calculating forward matrix elements of $B$ mesons, we want to make a few comments on this equation. Firstly, the decomposition of the operator $\mathcal P := \mathcal P_1 \mathcal P_2$ is of course not unique, and any other decomposition $\mathcal P = \mathcal P_1' \mathcal P_2'$ would have been good as well. Different decompositions will give estimates in terms of different lower-dimensional matrix elements and in the following we will always chose the decomposition in a way to obtain estimates in terms of the desired parameters.
Related to this is the position of the Dirac structure $\Gamma$ on the left-hand side of eq.~\eqref{eq:master1}. Since $\Gamma$ and $\mathcal P_i$ commute, we could have equally well placed it in the first matrix element with $\mathcal P_1$ or even split it up as $\Gamma:=\Gamma_1 \Gamma_2$.

Secondly, it is obvious from the derivation that this estimate can readily be generalized to two or more insertion of complete sets for estimating higher-dimensional matrix elements. We will see a case of this generalization at order $(\Lambda_{\rm QCD} / m_b)^5$ in sec.~\ref{sec:order5}. 

Finally, our last remark on eq.~\eqref{eq:master1} concerns the fact that it is only the tree-level approximation of the OPE, as stated in the derivation. Therefore, there are QCD corrections to our estimates coming from higher order terms in the OPE. However, these corrections can easily be included  by performing the OPE in eq.~\eqref{eq:OPE}  to higher order.  This point will be addressed in a subsequent publication.

\section{Contributions from lowest lying states} \label{sec:GSS}
The sums in~\eqref{master} and~\eqref{eq:master1} run over all intermediate states which have the appropriate 
quantum numbers. Obviously this sum cannot be performed analytically and one way of approaching this problem is to 
truncate the sum. In the following we truncate this sum after the lowest states that can contribute to the 
matrix element. These are either the ground states $Q$ and $Q^*$ (the pseudoscalar and vector meson
formed from the heavy quark $Q$ and a light antiquark) or the lowest lying states with angular momentum 
$\ell = 1$. 
Making use of heavy quark symmetries, these latter states consist of two degenerate doublets,
 for which  the spin of the light degrees of freedom is $j = 1/2$ and $j = 3/2$. 
As we will see, this will allow us to express the matrix elements up to order $(\Lambda_{\rm QCD} / m_b)^5$ in terms of just four parameters: the kinetic energy $\mu_\pi^2$, the chromomagnetic moment $\mu_G^2$ and the excitation energies $\epsilon_{1/2}$ and $\epsilon_{3/2}$ of the two $\ell=1$ doublets compared to the ground state (taking these last two is equivalent to using $\rho_D$ and $\rho_{LS}$). 

In order to implement spin symmetry, it is useful to define representation matrices for these states 
as
\begin{subequations} \label{eq:heavymesonrep}
\begin{align}
C (v) &= \sqrt{M_C} \frac{1+\fmslash{v}}{2} \gamma_5 
& J^P &=0^-, & j & = {1}/{2}, \label{eq:heavymesonrepC} \\    
C^* (v,\epsilon) & =  \sqrt{M_C} \frac{1+\fmslash{v}}{2} \fmslash{\epsilon} 
& J^P &=1^-, & j & = {1}/{2}, \label{eq:heavymesonrepCs} \\        
E(v) &= \sqrt{M_E} \frac{1+\fmslash{v}}{2}
& J^P &=0^+, & j & = {1}/{2}, \label{eq:heavymesonrepE} \\        
E^* (v,\epsilon) &=  \sqrt{M_E} \frac{1+\fmslash{v}}{2} \gamma_5 \fmslash{\epsilon} 
& J^P &=1^+, & j & = {1}/{2}, \label{eq:heavymesonrepEs} \\    
F^\mu (v,\epsilon) &= \sqrt{M_F} \sqrt{\frac{3}{2}} \frac{1+\fmslash{v}}{2} \gamma_5 
  \left[\epsilon^\mu - \frac{1}{3}\fmslash{\epsilon} (\gamma^\mu - v^\mu)\right] 
& J^P &=1^+, & j & = {3}/{2},  \label{eq:heavymesonrepF} \\    
F^{*\mu}  (v,\epsilon) &=  \sqrt{M_F}\sqrt{\frac{1}{2}} \frac{1+\fmslash{v}}{2} \epsilon^{\mu \nu} \gamma_\mu
& J^P &=2^+, & j & = {3}/{2}, \label{eq:heavymesonrepFs} \\       
G^\mu (v,\epsilon) &= \sqrt{M_G} \sqrt{\frac{3}{2}}\frac{1+\fmslash{v}}{2}  
  \left[\epsilon^\mu - \frac{1}{3}\fmslash{\epsilon} (\gamma^\mu + v^\mu)\right] 
& J^P &=1^-, & j & = {3}/{2}, \label{eq:heavymesonrepG}  \\    
G^{*\mu}  (v,\epsilon) &=  \sqrt{M_G} \sqrt{\frac{1}{2}}\frac{1+\fmslash{v}}{2} \gamma_5 \epsilon^{\mu \nu} \gamma_\mu
& J^P &=2^-, & j & = {3}/{2}\, , \label{eq:heavymesonrepGs}
\end{align}  
\end{subequations}
corresponding to the proper coupling of the light and heavy quark spins and the angular momentum~\cite{Falk:1990yz,Falk:1991nq}. 
Note that the states parametrized by 
$G^{(*)\mu}$ correspond to $\ell = 2$. They will not contribute in LLSA and are just given here for completeness. 
The polarization vectors $\epsilon^{(i)}_{\mu} $ and the traceless, 
symmetric polarization tensors $\epsilon^{(i)}_{\mu\nu}$ obey
\begin{subequations} \label{eq:polprop}
\begin{align}
v \cdot \epsilon^{(i)}& =0,	& \epsilon^{(i)} \cdot \epsilon^{(j)} & = -\delta^{ij}, & \sum_i \epsilon_\mu^{(i)} \epsilon_\nu^{(i)*} & = - g_{\mu\nu}^\perp, \label{eq:polprop1} \\
v^\mu \epsilon^{(i)}_{\mu\nu} & = 0, 	& \epsilon^{(i)}_{\mu\nu}\epsilon^{(j),\mu\nu} & = 2 \delta^{ij}, & \sum_i \epsilon^{(i)}_{\mu\nu}\epsilon^{(i)*}_{\alpha\beta} & = g^\perp_{\mu\alpha}g^\perp_{\nu\beta}+ g^\perp_{\mu\beta}g^\perp_{\nu\alpha}- \frac{2}{3}g^\perp_{\mu\nu}g^\perp_{\alpha\beta}.
\end{align}
\end{subequations}
These representations can be used to compute the matrix elements in~\eqref{master} and~\eqref{eq:master1}. 
Using the heavy  mass limit also for the $B$ meson, we obtain the ``trace formula'' \cite{Falk:1990yz,Falk:1991nq}
\begin{equation}
\braket{B | \bar{b} \mathcal P \Gamma Q  | n } = \Tr [ \bar{C} (v) \Gamma H (v) \mathcal H ],
\end{equation}
with $\bar C = \gamma^0 C^\dagger \gamma^0$ and $H(v)$ being the representations for the state $\ket{J^P,j}$ as given in eq.~\eqref{eq:heavymesonrep}.  
$\mathcal H$ represents the light degrees of freedom and depends on the derivatives contained in $\mathcal P$. 
The important feature, which reduces the number of independent coefficients and allows us to make meaningful predictions, is
that $\mathcal H$  is the same for each pair of doublets.
By definition of the heavy representations in eq.~\eqref{eq:heavymesonrep} the light degrees of freedom in $\mathcal H$ must conserve parity and as such they are composed of the  Dirac matrices 
$\mathds 1$, $\gamma^\mu_\perp$ and $i\sigma^{\mu\nu}_\perp$, together with the metric $g_{\mu \nu}$ and the $P$-even combination  $\epsilon_{\mu \nu \rho \sigma}\gamma^5$; the number of different independent combinations equals the number of parameters needed to describe theses matrix elements. The vector $v^\mu$ cannot appear as the indices appearing in $\mathcal H$ must all be perpendicular to $v^\mu$.

\subsection{Order $(\Lambda_{\rm QCD} / m_b)$ and $(\Lambda_{\rm QCD} / m_b)^2$}
We start with the simplest case, where $\mathcal P_1$ as well as $\mathcal P_2$ is only a single derivative. For illustration purposes we will go through the steps of the calculation in some detail. Starting from eq.~\eqref{eq:master1}, 
we get from the leading term of the $1/\omega$ expansion 
\begin{eqnarray}  \label{master3} 
&& \sum_n   (2 \pi)^3 \delta^3 (p_n^\perp) 
\Braket{ B(p_B) | \bar{b}_v   (iD_\mu^\perp)  Q_v | n } \, 
\Braket{ n |  \bar{Q}_v   (iD_\nu^\perp)  \Gamma b_v | B(p_B) }
\\ \nonumber && \qquad = 
\Braket{ B(p_B) | \bar{b}_v   (iD_\mu^\perp) (iD_\nu^\perp)    \frac{1+\fmslash{v}}{2}  \Gamma    
    b_v  | B(p_B) }.
\end{eqnarray} 
By rotational symmetry, the lowest-lying states that can contribute here are the two $\ell=1$ spin symmetry doublets~\eqref{eq:heavymesonrepE}-\eqref{eq:heavymesonrepEs}, from which only the 
two $1^+$-states yield a non-vanishing result. 

The representations of the light degrees of freedom carry one Lorentz index from the covariant derivative. In the case of the 
$j = 3/2$ doublet there is a second Lorentz index, which gets contracted with the one form the representation $F^\mu$ of the heavy mass state. 
Hence the light degrees of freedom are parametrized by two parameters $R$ and $R'$ as 
\begin{align} \label{eq:ldofdim4}
{\mathcal E}^\mu =  R \, \gamma^\mu_\perp 
\qquad \text{ and } \qquad 
 {\mathcal F}^{\mu \nu}  =  R'\,  g_\perp^{\mu \nu}.  
\end{align} 
Note that in $ {\mathcal F}^{\mu \nu}$ no $i\sigma^{\mu\nu}_\perp$-term appears, as the polarization index for spin $j\ge\tfrac{3}{2}$ belongs to a Rarita-Schwinger object that obeys $\psi_\mu \gamma^\mu = 0$ (this can be explicitly checked using eq.~\eqref{eq:heavymesonrepF}-\eqref{eq:heavymesonrepGs}).
Inserting this into the trace formula, we get 
\begin{subequations}  \label{eq:BiDB}
\begin{align} 
\Braket{B| \bar{b}_v   iD_\mu^\perp \Gamma Q_v  | 1^+, \tfrac{1}{2}} & =  
\Tr [ \bar{C} (v) \Gamma  E^* (v,\epsilon) \mathcal E_\mu ],  \\ 
\Braket{B| \bar{b}_v   iD_\mu^\perp \Gamma Q_v  | 1^+, \tfrac{3}{2}} & =  
\Tr [ \bar{C} (v) \Gamma  F^\nu (v,\epsilon) \mathcal F_{\mu \nu} ] ,
\end{align}
\end{subequations} 
where we used the notation $\ket{J^P,j}$ for the (intermediate) $Q$ states. 
Evaluating the traces, the only non-vanishing matrix elements for $\Gamma = \mathds{1}$ are 
\begin{subequations} \label{eq:BiDGQ}
\begin{align}
\Braket{B| \bar{b}_v   iD_\mu^\perp Q_v  | 1^+, \tfrac{1}{2} } &= -2 \sqrt{M_B M_E} R \, \epsilon_\mu , \\
\Braket{ B | \bar{b}_v  iD_\mu^\perp  Q_v  | 1^+, \tfrac{3}{2}} &= -2\sqrt{\frac{2}{3}} \sqrt{M_B M_{F}} \,R'  \, \epsilon_\mu, \\
\intertext{and the ones containing $\Gamma= i\sigma^\perp_{\alpha\beta}$ are}
\Braket{B | \bar{b}_v   iD_\mu^\perp  i\sigma^\perp_{\alpha\beta}Q_v  | 1^+,\tfrac{1}{2} } &=  - 4  \sqrt{M_B M_{E}} \, R \,  \epsilon_{[\alpha}g^\perp_{\beta]\mu}  ,\\
\Braket{B | \bar{b}_v   iD_\mu^\perp  i\sigma^\perp_{\alpha\beta}Q_v  | 1^+,\tfrac{3}{2} } &=  2\sqrt{\frac{2}{3}}  \sqrt{M_B M_{F}} \, R' \,  \epsilon_{[\alpha}g^\perp_{\beta]\mu} ,
\end{align}
\end{subequations}
where the square brackets $[\alpha\beta]$ denote antisymmetrization, i.e. $T_{[\alpha\beta]}= \frac{1}{2}(T_{\alpha\beta}-T_{\beta\alpha})$.
The key point of the ``lowest-lying  state saturation ansatz'' (LLSA) is to saturate the sum over all intermediate states by the lowest lying states only, which 
amounts here to the replacement 
\begin{equation} \label{eq:reducedsum}
\sum_n \ket{n} \bra{n} = \sum_{Pol} \int 
\frac{d^3p}{(2\pi)^3 2 E_p}
\left[ \ket{1^+, \tfrac{1}{2} } \bra{1^+, \tfrac{1}{2}}  
   +  \ket{1^+, \tfrac{3}{2} } \bra{1^+, \tfrac{3}{2}}  \vphantom{\int} \right]  + \cdots
\end{equation}  
on the left-hand side of~\eqref{master3}. The ellipses denote the higher excited states, which we shall omit. 
Using the polarization sums from~\eqref{eq:polprop1} and integrating over the momentum of the intermediate 
state we obtain the estimates for the the right-hand side of~\eqref{master3}
\begin{subequations} \label{eq:BiDiDGB}
\begin{align}
\Braket{B| \bar b_v iD_\mu^\perp i D_\nu^\perp b_v|B} & =  2 M_B \left( - |R|^2 \,  - \frac{2}{3}  |R'|^2 \right) \, g^\perp_{\mu\nu},  \\
\Braket{B| \bar b_v iD_\mu^\perp i D_\nu^\perp i \sigma^\perp_{\alpha\beta}b_v|B} & =  2 M_B \left(2 |R|^2 \,  - \frac{2}{3}  |R'|^2 \right) \, g^\perp_{\mu[\alpha}g^\perp_{\beta]\nu} .
\end{align}
\end{subequations}
For some further details on the derivation of eq.~\eqref{eq:BiDiDGB}  see Appendix~\ref{app:detailsdim4}.
These equations now allow us to relate $R$ and $R'$ to the kinetic energy $\mu_\pi^2$ and the chromomagnetic moment $\mu_G^2$ defined by
\begin{subequations} \label{eq:defofmupimug}
\begin{align}
2 M_B \mu_\pi^2 & = - \Braket{ B | \bar{b}_v    iD_\mu^\perp iD_\nu^\perp  b_v | B } g_\perp^{\mu\nu}, \\
2 M_B \mu_G^2 & = - \Braket{ B | \bar{b}_v    iD_\mu^\perp iD_\nu^\perp  i\sigma_\perp^{\mu\nu}  b_v | B }.
\end{align}
\end{subequations}
So  we finally obtain in the LLSA approximation
\begin{subequations} \label{eq:eq1}
\begin{align} 
9 |R|^2 & = \mu_\pi^2 - \mu_G^2  \\
6 |R'|^2 & = 2 \mu_\pi^2 + \mu_G^2 \, , 
\end{align} 
\end{subequations}
which reproduces the result of \cite{Mannel:2010wj} that the combination $\mu_\pi^2 - \mu_G^2$ only receives 
contributions of the $j = 1/2$ spin-symmetry doublet, while the combination $2 \mu_\pi^2 + \mu_G^2$ is fed 
from the $j = 3/2$ states.  In the calculations in the next subsections we will use eq.~\eqref{eq:eq1} to replace the parameters $R$ and $R'$ by $\mu_\pi^2$ and $\mu_G^2$.

\subsection{Order $(\Lambda_{\rm QCD} / m_b)^3$}
At  order $(\Lambda_{\rm QCD} / m_b)^3$ we have the Darwin term $\rho_D$ and the spin-orbit coupling $\rho_{LS}$, defined by
\begin{subequations}
\begin{align}
2 M_B \rho_D^3 & = \frac{1}{2} \Braket{ B | \bar{b}_v   \Big[ iD_\mu^\perp  ,\left[  iv\cdot D,  iD_\nu^\perp\right]\Big]  b_v | B }g_\perp^{\mu\nu}, \\
2 M_B \rho_{LS}^3 & = -\frac{1}{2} \Braket{ B | \bar{b}_v   \Big\{ iD_\mu^\perp  ,\left[  iv\cdot D,  iD_\nu^\perp \right]\Big\} i \sigma_\perp^{\mu\nu}  b_v | B } .
\end{align}
\end{subequations}
Hence we only have to  consider the matrix elements 
\begin{equation}
\Braket{B | \bar{b}_v iD_\mu^\perp(iv\cdot D)iD_\nu^\perp \Gamma b_v | B} ,
\end{equation} 
since the terms with $(iv\cdot D)$ on the very right (or left) must vanish due to the equations of motion. This means that we  have to consider the $1/\omega$-term  in our master equation~\eqref{eq:master1},
\begin{eqnarray}  \label{eq:eq314}
&& 
\sum_n   (2 \pi)^3 \delta^3 (p_n^\perp)  (-\epsilon_n) 
\Braket{ B(p_B) | \bar{b}_v  iD_\mu^\perp  Q_v  | n } \, \Braket{ n |  \bar{Q}_v    iD_\nu^\perp  \Gamma b_v  | B(p_B) }
\\ \nonumber && \qquad \qquad 
=  \Braket{ B(p_B) | \bar{b}_v    iD_\mu^\perp  (iv\cdot D)   \frac{1+\fmslash{v}}{2} iD_\nu^\perp  \Gamma    
    b_v    | B(p_B) }.
\end{eqnarray} 
According to the LLSA, we again saturate the sum on the left-hand side by the two $\ell=1$ 
spin symmetry doublets. Thus we pick up two new parameters, which we choose to be $\epsilon_{1/2}$ and $\epsilon_{3/2}$, 
 the excitation energies of the two spin symmetry doublets, instead of $\rho_D$ and $\rho_{LS}$.  
The matrix elements that appear in this approximation are given in eq.~\eqref{eq:BiDGQ}. Thus  we obtain
\begin{subequations} \label{eq:matrixmasterorder3}
\begin{align}
\Braket{B| \bar b_v iD_\mu^\perp (iv\cdot D) i D_\nu^\perp b_v|B} & =  2 M_B \left(  \epsilon_{1/2} |R|^2 \,  +  \frac{2}{3} \epsilon_{3/2} |R'|^2 \right) \, g^\perp_{\mu\nu} , \\
\Braket{B| \bar b_v iD_\mu^\perp (iv \cdot D) i D_\nu^\perp i \sigma^\perp_{\alpha\beta}b_v|B} & =  2 M_B \left(-2 \epsilon_{1/2} |R|^2 \,  +\frac{2}{3} \epsilon_{3/2} |R'|^2 \right) \, g^\perp_{\mu[\alpha}g^\perp_{\beta]\nu} .
\end{align}
\end{subequations}
From these we can  eliminate $R$ and $R'$ using eq.~\eqref{eq:eq1}. Plugging this into the definition of $\rho_D$ and $\rho_{LS}$, this then yields
\begin{subequations} \label{eq:resultsforrho}
\begin{align}
\rho_D^3  		& = \frac{1}{3} \epsilon_{1/2}(\mu_\pi^2-\mu_G^2)+ \frac{1}{3}\epsilon_{3/2}(2\mu_\pi^2 + \mu_G^2),  \\
\rho_{LS}^3  	& =  \frac{2}{3}  \epsilon_{1/2}(\mu_\pi^2-\mu_G^2)-\frac{1}{3}\epsilon_{3/2}(2\mu_\pi^2 + \mu_G^2).
\end{align}
\end{subequations}
Again we observe that the combination $ \rho_D^3 + \rho_{LS}^3$ only is driven by the $j=1/2$ intermediate states and likewise 
the $j = 3/2$ states determine the combination $ 2 \rho_D^3 - \rho_{LS}^3$.
The numerical values for these estimates are discussed in sec.~\ref{sec:numerics}.

\subsection{ Order $(\Lambda_{\rm QCD} / m_b)^4$ }
At order $(\Lambda_{\rm QCD} / m_b)^4$ we have nine independent matrix elements, four  spin-singlets and five spin-triplets. These are  parametrized by~\cite{Mannel:2010wj} 
\begin{subequations} \label{eq:defofms}
\begin{align}
2M_B m_1 & = \Braket{B| \bar b_v iD^\perp_\mu iD^\perp_\nu iD^\perp_\rho iD^\perp_\sigma b_v |B} \frac{1}{3}(g_\perp^{\mu\nu}g_\perp^{\rho\sigma}+g_\perp^{\mu\rho}g_\perp^{\nu\sigma}+g_\perp^{\mu\sigma}g_\perp^{\nu\rho}) , \\
2M_B m_2 & = \Braket{B| \bar b_v [iD^\perp_\mu,iv\cdot D][i v \cdot D, iD^\perp_\sigma] b_v |B} g_\perp^{\mu\sigma} , \\
2M_B m_3 & = \Braket{B| \bar b_v [iD^\perp_\mu,iD^\perp_\nu][iD^\perp_\rho, iD^\perp_\sigma] b_v |B} g_\perp^{\mu\rho} g_\perp^{\nu\sigma} , \\
2M_B m_4 & = \Braket{B| \bar b_v \Big\{ iD^\perp_\mu,\big[iD^\perp_\nu,[iD^\perp_\rho, iD^\perp_\sigma ] \big] \Big\} b_v |B} g_\perp^{\nu\rho} g_\perp^{\mu\sigma} , \\
2M_B m_5 & = -\Braket{B| \bar b_v [iD^\perp_\mu,iv\cdot D][i v \cdot D, iD^\perp_\sigma] i\sigma_\perp^{\mu\sigma}b_v |B}  , \\
2M_B m_6 & = -\Braket{B| \bar b_v [iD^\perp_\mu,iD^\perp_\nu][iD^\perp_\rho, iD^\perp_\sigma] i \sigma_\perp^{\nu\rho} b_v |B}  g_\perp^{\mu\sigma} , \\
2M_B m_7 & = -\Braket{B| \bar b_v \Big\{ \{ iD^\perp_\mu,iD^\perp_\nu\}, [iD^\perp_\rho, iD^\perp_\sigma] \Big\} i \sigma_\perp^{\mu\sigma} b_v |B}  g_\perp^{\nu\rho} , \\
2M_B m_8 & = -\Braket{B| \bar b_v \Big\{ \{ iD^\perp_\mu,iD^\perp_\nu\}, [iD^\perp_\rho, iD^\perp_\sigma] \Big\} i \sigma_\perp^{\rho\sigma} b_v |B}  g_\perp^{\mu\nu} , \\
2M_B m_9 & = -\Braket{B| \bar b_v \Big[ iD^\perp_\mu,\big[iD^\perp_\nu,[iD^\perp_\rho, iD^\perp_\sigma ] \big] \Big] i \sigma_\perp^{\rho\mu} b_v |B} g_\perp^{\nu\sigma}.
\end{align}
\end{subequations}
Two of these matrix elements, $m_2$ and $m_5$, contain time derivatives and are obtained  by the $k=2$ term in eq.~\eqref{eq:master1} analogously to $\rho_D$ and $\rho_{LS}$
\begin{subequations} \label{eq:matrixmasterorder4a}
\begin{align} 
m_2 & = -\frac{1}{3} \left( \epsilon_{1/2}^2(\mu_\pi^2-\mu_G^2)+\epsilon_{3/2}^2(2\mu_\pi^2 + \mu_G^2) \right), \\
m_5 & = \frac{1}{3}\left( 2 \epsilon_{1/2}^2(\mu_\pi^2-\mu_G^2)+\epsilon_{3/2}^2(2\mu_\pi^2 + \mu_G^2)\right).
\end{align}
\end{subequations}
The other matrix elements contain only spatial derivatives. We insert the complete set of states in the middle, and keep only the contributions from the negative parity $j=1/2$ states, which are related to the matrix elements of the form $\Braket{B| iD^\perp iD^\perp \Gamma |B}$ given in  in eq.~\eqref{eq:defofmupimug}, so the estimates will  contain only $\mu_\pi^2$ and $\mu_G^2$. The contributions from the  negative parity $j=3/2$ states, which would introduce new parameters, are not kept in LLSA. The complete list of non-vanishing matrix elements also containing these states are given in app.~\ref{app:detailsdim5}.
The relevant light degrees of freedom are therefore parametrized by
\begin{align} \label{eq:dim5brownmuck}
{\mathcal C}_{\mu\nu} &=  \frac{1}{3} \mu_\pi^2  g^\perp_{\mu\nu} - \frac{1}{6} \mu_G^2 i\sigma^\perp_{\mu\nu},
\end{align} 
which leads to
\begin{subequations} \label{eq:BiDiDiDiDB}
\begin{align}
\Braket{B | \bar{b}_v   iD_\mu^\perp iD_\nu^\perp iD_\rho^\perp iD_\sigma^\perp  b_v  | B } 
& = 2 M_B \frac{1}{18} \left( 2  \mu_\pi^4 g^\perp_{\mu\nu}g^\perp_{\rho\sigma}  -  \mu_G^4  g^\perp_{\mu[\rho} g^\perp_{\sigma]\nu} \right),
\\
\Braket{B | \bar{b}_v   iD_\mu^\perp iD_\nu^\perp iD_\rho^\perp iD_\sigma^\perp  i \sigma^\perp_{\alpha\beta}b_v  | B } 
& =  2 M_B \frac{\mu_G^2}{9} \Big\{ \mu_\pi^2   \left(g^\perp_{\mu[\alpha}g^\perp_{\beta]\nu} g^\perp_{\rho\sigma} + g^\perp_{\mu\nu}   g^\perp_{\rho[\alpha} g^\perp_{\beta]\sigma} \right)  \\
		 & \qquad \qquad \qquad  +  2 \mu_G^2  \Big[ \left[g^\perp_{\mu[\alpha} g^\perp_{\beta]\rho}g^\perp_{\nu\sigma}\right]_{\mu\nu}  \Big]_{\rho\sigma}  \Big\}. \nonumber
\end{align}
\end{subequations}
Using the definitions~\eqref{eq:defofms}, the parameters $m_i$  can then easily be calculated. The results are shown in tab.~\ref{tab:resultsform} (for the numerical values  in this table see the discussion in sec.~\ref{sec:numerics}).
\begin{table} 
{\small 
\noindent
\tabulinesep = 1mm
\begin{tabu} to 1.00\textwidth {||c|X[c]|c||c|X[c]|c|} \hline\hline
& Expression & $\!\! \!\scalebox{.85}{}\!\!$ & & Expression & $\!\! \!\scalebox{.85}{}\!\!$  
\\ \hline\hline
 $m_1$ & $ \frac{5}{9}\mu_\pi^4 $ & $9.5$ &
$m_2$ & $ -\frac{\epsilon_{1/2}^2}{3} (\mu_\pi^2-\mu_G^2) - \frac{\epsilon_{3/2}^2}{3}(2\mu_\pi^2 + \mu_G^2)	 $ & $-8.2$ 
 \\ \hline 
$m_3$ & $  -\frac{2}{3}\mu_G^4 $ & $ -7.7 $ \rule[-5pt]{0pt}{17pt} &
$m_4$ & $ \!\!   \mu_G^4+ \frac{4}{3}\mu_\pi^4   \!\! $ & $  34.4 $ 
 \\ \hline 
$m_5$ & $  -\frac{2\epsilon_{1/2}^2}{3} (\mu_\pi^2-\mu_G^2) + \frac{\epsilon_{3/2}^2}{3}(2\mu_\pi^2 + \mu_G^2) $ & $ 7.0 $ &
$m_6$ & $   \frac{2}{3}\mu_G^4$ & $  7.7  $\rule[-5pt]{0pt}{17pt}
 \\ \hline
$m_7$ & $ -\frac{8}{3} \mu_\pi^2 \mu_G^2 $ & $ -37.5 $ &
$m_8$ & $ -8 \mu_\pi^2 \mu_G^2 $ & $ -112.6 $ 
 \\ \hline 
$m_9$ & $  \!\!  \mu_G^4 - \frac{10}{3} \mu_\pi^2 \mu_G^2  \!\! $ &  $ -35.4 $ \rule[-5pt]{0pt}{17pt}
	&  & & \\ 
\hline\hline
\end{tabu}
}
\caption{Expressions and values for the dimension seven matrix elements $m_i$. The numerical values are in units of $10^{-2} \text{ GeV}^4$.} 
\label{tab:resultsform} 
\end{table}

\subsection{Order $(\Lambda_{\rm QCD} / m_b)^5$ \label{sec:order5}}
At order $(\Lambda_{\rm QCD} / m_b)^5$ the number of independent matrix elements proliferates even more, resulting in seven spin-singlet and eleven spin-triplet operators. We chose to define these eighteen parameters according to~\cite{Mannel:2010wj} 
\begin{subequations}
\begin{align}
2M_B r_1 &= \Braket{ B | \bar b_v \,i  D_\mu^\perp\, (i v \cdot D)^3\, i  D^\mu_\perp \, b_v | B } , \\
2M_B r_2 &= \Braket{ B | \bar b_v \,i  D_\mu^\perp\, (i v \cdot D)\, i  D^\mu_\perp\, i  D_\nu^\perp\, i  D^\nu_\perp \, b_v | B }  , \\
2M_B r_3 &= \Braket{ B | \bar b_v \,i  D_\mu^\perp\, (i v \cdot D)\, i  D_\nu^\perp\, i D_\perp^\mu\, i  D_\perp^\nu \, b_v | B }  , \\  
2M_B r_4 &= \Braket{ B | \bar b_v \,i  D_\mu^\perp\, (i v \cdot D)\, i  D_\nu^\perp\, i D^\nu_\perp\, i  D^\mu_\perp \, b_v | B }  , \\  
2M_B r_5 &= \Braket{ B | \bar b_v \,i  D_\mu^\perp\, i  D^\mu_\perp\,(i v \cdot D)\,  i D_\nu^\perp\, i  D^\nu_\perp \, b_v | B } , \\  
2M_B r_6 &= \Braket{ B | \bar b_v \,i  D_\mu^\perp\, i  D_\nu^\perp\, (i v \cdot D)\, i D^\nu_\perp\, i  D^\mu_\perp \, b_v | B }  , \\  
2M_B r_7 &= \Braket{ B | \bar b_v \,i  D_\mu^\perp\, i  D_\nu^\perp\, (i v \cdot D)\, i D^\mu_\perp\, i  D^\nu_\perp \, b_v | B }  , \\  
2M_B r_{8} &= - \Braket{ B | \bar b_v \,i   D_\alpha^\perp \, (i v \cdot D)^3\, i   D_\beta^\perp \, i \sigma_\perp^{\alpha \beta }\,b_v | B }  , \\  
2M_B r_{9} &= - \Braket{ B | \bar b_v \,i   D_\alpha^\perp \, (i v \cdot D)\, i   D_\beta^\perp \, i   D_\mu^\perp\, i   D^\mu_\perp \,i \sigma_\perp^{\alpha \beta }\, b_v | B }  , \\  
2M_B r_{10} &= - \Braket{ B | \bar b_v \,i   D_\mu^\perp\, (i v \cdot D)\, i D^\mu_\perp\, i   D_\alpha^\perp \, i   D_\beta^\perp  \,i \sigma_\perp^{\alpha \beta }\, b_v | B }  , \\
2M_B r_{11} &= - \Braket{ B | \bar b_v \,i   D_\mu^\perp\, (i v \cdot D)\, i   D_\alpha^\perp \, i   D^\mu_\perp\, i   D_\beta^\perp  \,i \sigma_\perp^{\alpha \beta }\, b_v | B }  , \\  
2M_B r_{12} &= -\Braket{ B | \bar b_v \,i   D_\alpha^\perp \, (i v \cdot D)\, i D_\mu^\perp\, i   D_\beta^\perp \, i   D^\mu_\perp \,i \sigma_\perp^{\alpha \beta }\, b_v | B }  , \\
2M_B r_{13} &= - \Braket{ B | \bar b_v \,i   D_\mu^\perp\, (i v \cdot D)\, i   D_\alpha^\perp \, i   D_\beta^\perp \, i   D^\mu_\perp \,i \sigma_\perp^{\alpha \beta }\, b_v | B }  , \\  
2M_B r_{14} &= - \Braket{ B | \bar b_v \,i   D_\alpha^\perp \, (i v \cdot D)\, i D_\mu^\perp\, i   D^\mu_\perp\, i   D_\beta^\perp  \, i \sigma_\perp^{\alpha \beta }\, b_v | B }  , \\
2M_B r_{15} &= - \Braket{ B | \bar b_v \,i   D_\alpha^\perp \, i   D_\beta^\perp \, (i v \cdot
D)\, i   D_\mu^\perp\, i   D^\mu_\perp \, i \sigma^{\alpha \beta }\, b_v | B }  , \\  
2M_B r_{16} &= - \Braket{ B | \bar b_v \,i   D_\mu^\perp\, i   D_\alpha^\perp \, (i v \cdot
D)\, i   D_\beta^\perp \, i   D^\mu_\perp \, i \sigma_\perp^{\alpha \beta }\, b_v | B } , \\  
2M_B r_{17} &= - \Braket{ B | \bar b_v \,i   D_\alpha^\perp \, i   D_\mu^\perp\, (i v \cdot
D)\, i   D^\mu_\perp\, i   D_\beta^\perp  \, i \sigma_\perp^{\alpha \beta }\, b_v | B }  , \\  
2M_B r_{18} &= -\Braket{ B | \bar b_v \,i   D_\mu^\perp\, i   D_\alpha^\perp \, (i v \cdot D)\, i   D^\mu_\perp\, i   D_\beta^\perp  \, i \sigma_\perp^{\alpha \beta }\, b_v | B } .
\end{align}
\end{subequations}
As for the $m_i$ parameters 
these different matrix elements have to be handled in slightly different ways to obtain a result in LLSA. 
For details of these calculations see  appendices~\ref{app:detailsdim5} and~\ref{app:detailsdim6}.
\begin{itemize}
\item $\mathbf {r_{1,8}}$ : In complete analogy to $\rho_{D,LS}$ and $m_{2,5}$ respectively, these two parameters can be obtained from
	\begin{subequations} \label{eq:matrixmasterorder5a}
	\begin{align}
\Braket{B| \bar b_v iD_\mu^\perp (iv D)^3 i D_\nu^\perp b_v|B} & =  2 M_B \left(  \epsilon_{1/2}^3 |R|^2 \,  +  \frac{2}{3} \epsilon_{3/2}^3 |R'|^2 \right) \, g^\perp_{\mu\nu},  \\
\Braket{B| \bar b_v iD_\mu^\perp (iv D)^3 i D_\nu^\perp i \sigma^\perp_{\alpha\beta}b_v|B} & =  2 M_B \left(-2 \epsilon_{1/2}^3 |R|^2 \,  +\frac{2}{3} \epsilon_{3/2}^3 |R'|^2 \right) \, g^\perp_{\mu[\alpha}g^\perp_{\beta]\nu} .
	\end{align}
	\end{subequations}
\item $\mathbf {r_{2-4,9-14}}$: We perform the insertion of eq.~\eqref{eq:master1} between the second and third space derivative. By rotational symmetry only the states with an even $\ell$ can contribute. Thus in LLSA we only keep the contributions from the two $\ell=0$ states, $0^-$ and $1^-$, which only contain $\mu_\pi$, $\mu_G$, $\rho_D$ and $\rho_{LS}$. The resulting uncontracted matrix elements are
	\begin{subequations} \label{eq:matrixmasterorder5b}
	\begin{align}
\Braket{B | \bar{b}_v   iD_\mu^\perp (iv\cdot D) iD_\nu^\perp iD_\rho^\perp iD_\sigma^\perp  b_v  | B } 
& = 2 M_B \frac{1}{18} \left( -2  \mu_\pi^2 \rho_D^3 g^\perp_{\mu\nu}g^\perp_{\rho\sigma}  -  \mu_G^2 \rho_{LS}^3  g^\perp_{\mu[\rho} g^\perp_{\sigma]\nu} \right),
\\
\Braket{B | \bar{b}_v   iD_\mu^\perp(iv\cdot D) iD_\nu^\perp iD_\rho^\perp iD_\sigma^\perp  i \sigma^\perp_{\alpha\beta}b_v  | B } 
& =  2 M_B \frac{1}{9} \Big\{ \mu_\pi^2 \rho_{LS}^3  g^\perp_{\mu[\alpha}g^\perp_{\beta]\nu} g^\perp_{\rho\sigma} - \mu_G^2 \rho_D^3    g^\perp_{\mu\nu}   g^\perp_{\rho[\alpha} g^\perp_{\beta]\sigma}   \nonumber \\
		 & \qquad    +  2 \mu_G^2 \rho_{LS}^3 \Big[ \left[g^\perp_{\mu[\alpha} g^\perp_{\beta]\rho}g^\perp_{\nu\sigma}\right]_{\mu\nu}  \Big]_{\rho\sigma}  \Big\} .
	\end{align}
	\end{subequations}
Note the analogy of these with eq.~\eqref{eq:BiDiDiDiDB}.
\item $\mathbf {r_{5-7,15-18}}$: There are two different ways, one can approach these matrix elements. One way is to make the insertion in the middle analogous to eq.~\eqref{eq:eq314}, corresponding to taking the $1/\omega$ piece in eq.~\eqref{eq:master1}. But, as mentioned above, in LLSA only the ground states contribute to these matrix elements and these have excitation energy zero. Hence we  obtain zero for all these matrix elements in this approximation.

Another way is to notice that the derivation of eq.~\eqref{eq:master1} is easy to generalize to the case of multiple insertions of complete states and try to get an estimate for $r_{5-7,15-18}$ in LLSA this way. 
One can perform two insertions, one after the first space derivative and one before the last space derivative. This will lead to the matrix elements of eq.~\eqref{eq:BiDGQ} and matrix elements of the form $\Braket{1^+,j|iD^\perp(iv\cdot D)iD^\perp|1^+,j'}$. These latter matrix elements contain new parameters, which have not been accounted for in our preceding analysis, as they do not contribute in LLSA. 

However, these matrix elements mentioned above look similar to the ones appearing in $\rho_D$ and $\rho_{LS}$. Thus, as a consistency check, we have verified, that  if one performed a double insertion and assumed the new parameters to be of order $\rho_D$ and $\rho_{LS}$, the matrix elements $r_{5-7,15-18}$ are indeed numerically smaller than the others. 
This is consistent with the vanishing result from the first approach to these matrix elements.
\end{itemize}
Using eq.~\eqref{eq:matrixmasterorder5a} and~\eqref{eq:matrixmasterorder5b} we get the LLSA estimates for the coefficients $r_i$, which are given in tab.~\ref{tab:resultsforr} (for numerics see again sec.~\ref{sec:numerics}). 
\begin{table} 
{\small 
\noindent
\tabulinesep = 1mm
\begin{tabu} to 1.00\textwidth {||l|c|X[c]||l|c|X[c]||} \hline\hline
   & Expression & $\!\!\! \!\scalebox{.85}{}\!\!\!$   & &
Expression &  $\!\!\!\!\scalebox{.85}{}\!\!\!$  \\ \hline\hline
$\!r_1\!$ & $ \frac{\epsilon_{1/2}^2}{3}(\rho_D^3+\rho_{LS}^3) + \frac{\epsilon_{3/2}^2}{3}(2\rho_D^3-\rho_{LS}^3)   $ & $   3.6$ &
$\!r_2\!$ & $  - \mu_\pi^2 \rho_D^3   $ & $  -7.6   $
 \\ \hline
$\!r_3\!$ & $ -\frac{1}{6}\mu_G^2 \rho_{LS}^3 - \frac{1}{3} \mu_\pi^2 \rho_D^3  $ & $   -1.7  $ \rule[-5pt]{0pt}{17pt} &
$\!r_4\!$ & $ \frac{1}{6}\mu_G^2 \rho_{LS}^3 - \frac{1}{3} \mu_\pi^2 \rho_D^3  $ & $ -3.4   $  \rule[-5pt]{0pt}{17pt}
 \\ \hline
$\!r_5\!$ & $ \!\! 0 \!\!   $ & $ 0   $ &
$\!r_6\!$ & $ \!\! 0 \!\!     $ & $ 0  $
 \\ \hline
$\!r_7\!$ & $ 	\!\! 0 \!\!    $ & $  0  $ &
$\!r_8\!$ & $ \frac{2\epsilon_{1/2}^2}{3}(\rho_D^3+\rho_{LS}^3) - \frac{\epsilon_{3/2}^2}{3}(2\rho_D^3-\rho_{LS}^3)   $ & $   -3.2 $ 
 \\ \hline
$\!r_9\!$ & $ -\mu_\pi^2 \rho_{LS}^3    $ & $ 6.4  $ &
$\!r_{10}\!$ & $  \mu_G^2 \rho_D^3   $ & $  6.2   $
 \\ \hline
$\!r_{11}\!$ & $  \frac{1}{6} (2\mu_\pi^2 - \mu_G^2) \rho_{LS}^3 + \frac{1}{3} \mu_G^2 \rho_D^3	  $ & $  0.8   $ &
$\!r_{12}\!$ & $   - \frac{1}{6} (2\mu_\pi^2+\mu_G^2) \rho_{LS}^3 - \frac{1}{3} \mu_G^2 \rho_D^3   $ & $ 0.9   $ 
 \\ \hline
$\!r_{13\!}$ & $ \frac{1}{6} (2\mu_\pi^2 + \mu_G^2) \rho_{LS}^3 - \frac{1}{3} \mu_G^2 \rho_D^3 $ & $ -5.1  $ &
$\!r_{14}\!$ & $ -\frac{1}{6} (2\mu_\pi^2-\mu_G^2) \rho_{LS}^3 + \frac{1}{3} \mu_G^2 \rho_D^3 \!\!\! $ & $ 3.3 $ 
 \\ \hline
$\!r_{15}\!$ & $  \!\! 0 \!\!    $ & $ 0  $ &
$\!r_{16\!}$ & $ \!\! 0 \!\!    $ & $  0  $
 \\ \hline
$\!r_{17}\!$ & $  \!\! 0 \!\!   $ & $ 0   $ &
$\!r_{18}\!$ & $ \!\! 0 \!\!    $ & $  0$ 
 \\ 
\hline\hline
\end{tabu}
}
\caption{Expressions and values for the dimension eight matrix elements $r_i$. The numerical values are given in units of $10^{-2} \text{ GeV}^5$. The parameters $\rho_D$ and $\rho_{LS}$ are given in~\eqref{eq:resultsforrho}. } 
\label{tab:resultsforr} 
\end{table} 

\subsection{Results and numerical estimates\label{sec:numerics}}
In the previous subsections we have expressed all higher order matrix elements up to order $(\Lambda_{\rm QCD} / m_b)^5$ in terms of the four known 
parameters $\mu_\pi$, $\mu_G$, $\epsilon_{1/2}$ and $\epsilon_{3/2}$ (we just chose to express some of the formulae in terms of $\rho_D$ and $\rho_{LS}$ for compacter notation, but these can be replaced using eq.~\eqref{eq:resultsforrho}). 
 
To get a feeling for the size of these parameters, we use the values~\cite{Beringer:1900zz, Gambino:2013rza, Acciarri:1999jx, Aaltonen:2008aa,Aaltonen:2013atp}
\begin{equation} \label{eq:input}
\mu_\pi^2 = 0.414 \text{ GeV}^2, \quad \mu_G^2 = 0.340 \text{ GeV}^2, \quad \epsilon_{1/2}=0.390 \text{ GeV},\quad \epsilon_{3/2}=0.476 \text{ GeV}
\end{equation}
to obtain numerical values for our estimates. For a comment on the errors of these input parameters see below at the end of this section.

 First, we see that the numerical values for the Darwin and spin-orbit coupling are given  by
\begin{align}
\rho_D^3 = 0.21 \text{ GeV}^3
\qquad \text{ and } \qquad
\rho_{LS}^3 = -0.17 \text{ GeV}^3.
\end{align}
Comparing these to the values fitted to experiment~\cite{Gambino:2013rza}, given by $\rho_D^3 =( 0.154 \pm 0.045 )\text{ GeV}^3$ and $\rho_{LS}^3 = (-0.147 \pm 0.098) \text{ GeV}^3$, we see that our estimates are in very good agreement and yield consistent results. The numerical values for the higher order parameters  $m_i$ and $r_i$ with  these input parameters are shown in tables~\ref{tab:resultsform} and~\ref{tab:resultsforr}. 

Of course these estimates are not very precise due to the truncation of the sum  in eq.~\eqref{eq:master1}, but  should be regarded as good ballpark estimates. We will give an approximation and discussion of the systematical error from the truncation of the sum in the next section. As we will see, this error is comparatively large, which is also the reason why we have refrained from giving the errors of our input parameters in eq.~\eqref{eq:input} and showing their impact on the numerical values of the matrix element estimates.

However, despite the sizable errors, the use of our estimate is twofold. First of all, we expect to have the correct signs and also the correct relative sizes of the matrix elements.   Secondly, we also expect to have 
the proper correlations between the matrix elements.  As one example for this, one can observe that we obtained $m_3 = - m_6$. While the precise factors  will very likely not stand up to scrutiny, their  order of magnitude and sign should.

\section{Estimate of the uncertainties} \label{sec:uncertainties}
Although our approach is quite systematic, an estimate of the uncertainties is not easy. The quality of the 
uncertainties is very different: while the uncertainties in $\mu_\pi^2$, $\mu_G^2$, $\epsilon_{1/2}$ and $\epsilon_{3/2}$ 
and the ones induced by QCD corrections are almost trivial to discuss, the uncertainty 
induced by the  truncation of the sum over intermediate states is very difficult to estimate. 
Clearly  a reliable estimate of this uncertainty  would require a non-perturbative solution of QCD. To this end, 
we have to rely on simple estimates based on toy models. While this will not give us a very robust estimate of the error, 
we will at least  get some insight, how far we can trust the result from the truncated series. 

The left-hand side of the master formula~\eqref{master} can be written as a dispersion integral over a spectral 
function $\rho(\omega)$ 
\begin{equation}
\Delta (\omega) = \int \frac{d\omega'}{2 \pi} \frac{\rho(\omega')}{\omega - \omega'},
\end{equation} 
with 
\begin{equation}
\rho (\omega) = \sum_n  (2 \pi)^3 \delta^3 (p_n^\perp) \delta(\omega-\epsilon_n) 
\Braket{ B(p_B) | \bar{b}_v   \mathcal P_1 Q_v | n } \, \Braket{ n |  \bar{Q}_v   \mathcal P_2  \Gamma b_v   | B(p_B) }.
\end{equation}
In order to discuss the effect of truncation, we strongly simplify the spectral function and use as a toy model 
an ansatz which has been discussed by Shifman to investigate duality violations~\cite{Blok:1997hs,Shifman:2000jv,Shifman:2003de}.  In this toy model, 
the spectral function consists of infinitely many, equally spaced narrow resonances, hence 
$\epsilon_n = n \Lambda$ and thus 
\begin{equation}
\tilde{\rho} (\omega) = \sum_n  g(n)  \delta(\omega - n \Lambda). 
\end{equation}
Inserting this into the dispersion relation, we get for this toy model 
\begin{equation} \label{Del1}
\tilde{\Delta} (\omega) = \frac{1}{2 \pi} \sum_n g(n) \frac{1}{\omega - n \Lambda}
\end{equation} 
The factor $g(n)$ takes into account the decrease of the matrix elements with increasing excitation quantum 
number $n$. In order to estimate this, we make use of a non-relativistic model for the heavy mesons with an hard-wall 
spherical box potential. The solution of the 
Schr\"odinger equation for the radial wave functions are the spherical Bessel functions, and for example the matrix elements 
that appear in  eq.~\eqref{eq:BiDB}  obey
\begin{equation}
\braket{\ell=0| \vec \nabla | \ell=1,m_z,n} \propto \frac{1}{n} \vec e_{m_z} +\dots,
\end{equation} 
where $\vec e_{m_z}$ denotes the polarization vectors for $m_z=\pm 1,0$. 
We take this as a general feature which we assume to be true also for the real QCD case: The radially excited 
states have in their radial wave function $n$ nodes, where $n$ is the quantum number for the radial excitations. 
Each node involves a sign change of the radial wave function, which results in an increasingly smaller overlap of the 
radially excited states with the ground state; in the non-relativistic model this scales as $1/n$. 

Assuming this, we set
\begin{equation}
g(n) = g_0 \frac{1}{n^2}, 
\end{equation} 
in which case the summation in~\eqref{Del1} can be performed, and yields
\begin{equation}
\tilde{\Delta} (\omega) = \frac{g_0}{2 \pi \Lambda} \frac{1}{x^2} \left[ 
\gamma + \psi (1-x)  + \frac{\pi^2}{6} x  \right],
\end{equation} 
where $x = \omega/\Lambda$ and $\psi (z)$ is the derivative of the logarithm of Euler's Gamma function.  
As discussed above, we perform an expansion for large (negative) $\omega$, and the asymptotic form of 
$\tilde\Delta (\omega)$ is given by 
\begin{equation}\label{Res1}
 \tilde{\Delta} (\omega) \longrightarrow \frac{g_0}{2 \pi \Lambda} \left[\frac{\pi^2}{6}\right]  \left(\frac{1}{x} \right)  
 \quad \mbox {as} \quad  x \to - \infty.
\end{equation} 
This result has to be compared to the one obtained form the truncation of the series after the first term. 
Taking only the fist term in our toy model, we get 
\begin{equation} \label{Res2}
 \tilde{\Delta}^{(1)}  (\omega) = \frac{g_0}{2 \pi \Lambda} \frac{1}{x-1} 
 \longrightarrow \frac{g_0}{2 \pi \Lambda} \left(\frac{1}{x} \right)   \quad \mbox {as} \quad  x \to - \infty.
\end{equation} 
Comparing eq.~\eqref{Res1} with eq.~\eqref{Res2}, we see that the relative uncertainty from omitting  the higher order terms in the series is given by 
\begin{equation} \label{est1} 
 \left[\frac{\pi^2}{6}\right]  - 1 \, \,\sim  64\%. 
\end{equation} 

Obviously this result strongly depends on the function $g(n)$, and the non-relativistic reasoning may fail. 
One can go through the same steps and assume ad hoc a different power dependence for $g(n)$, like $g(n) = g_0/n^{3}$ in which case~\eqref{est1} becomes $\zeta(3)  - 1\, \,  \sim 20\%  $. With higher powers of $1/n$ the uncertainty in the toy model truncation becomes smaller, supporting the intuitive picture that the overlap of the ground state wave function with excited state wave functions with an increasing number of nodes becomes more and more negligible and thus leads to better results for the truncation.

The toy model calculations as well as comparisons to simple calculations for different non-relativistic quantum systems all indicate  that the uncertainty due to the truncation of the series at lowest order is roughly of the order of 50\%. On the first glance this might sound terrible, as it will not allow for a precise prediction unless higher excited states are included (e.g. the $\ell=2$ states~\eqref{eq:heavymesonrepG} and \eqref{eq:heavymesonrepGs}). However, these higher states can be included systematically with more parameters and will then result in  more precise estimates.

\section{Summary \label{sec:conclusion}} 
We have systematically derived equation~\eqref{eq:master1} to express matrix elements in HQET by a sum of products of lower-dimensional matrix elements, in analogy of an insertion to  a complete set of states in non-relativistic quantum mechanics. As given, eq.~\eqref{eq:master1} is the tree-level term of an OPE, but it can be generalized systematically  to include higher-order QCD corrections. 

Furthermore, we have explicitly shown how this equation can be used to derive estimates for $B$ meson matrix elements up to order $(\Lambda_{\rm QCD} / m_b)^5$. In our ansatz we have only kept the lowest contributing states and express all the matrix elements in terms of just four parameters, the kinetic energy $\mu_\pi^2$, the chromo-magnetic moment $\mu_G^2$ and the excitation energies of the lowest contributing states $\epsilon_{1/2}$ and $\epsilon_{3/2}$.

To estimate the error of our estimates that is due to the truncation of the series, we made use of toy models for the spectral function and of comparisons to non-relativistic quantum mechanics. This leads to an estimate of the error of $\sim 50 \%$, when only including the lowest lying states in the sum. Of course this error will be dramatically reduced when higher excitations are included. Furthermore, even with a relatively large error, our estimates yield correlations amongst different matrix elements and allow for order of magnitude estimates and the determination of their signs.

\subsection*{Acknowledgements}
This work was supported by the German Research Foundation DFG within the Research Unit FOR 1873: {\it Quark Flavour Physic and Effective Field Theories}. 
We would like to thank Sascha Turczyk  for pointing out a sign error in section~\ref{sec:GSS} of the first version of this paper.

\begin{appendix}

\section{Some details of the calculation} \label{app:details}
In this appendix we collect some of the details for the calculation of the matrix elements estimates. In particular we will state the form of the light degrees of freedom and the non-vanishing matrix elements that we used to get the results of sec.~\ref{sec:GSS}. We will give the results for all four doublets appearing in eq.~\eqref{eq:heavymesonrep} even though we are only interested in LLSA, so there is a proliferation of parameters in this appendix, which do not appear in the main body of this paper.

\subsection{Dimension 4 matrix elements \label{app:detailsdim4}}
The matrix elements at dimension four are of the form $\Braket{B|\bar b_v iD^\perp \Gamma Q_v |n}$. The light degrees of freedom for the four $j=1/2$ and $j=3/2$ doublets are given as
\begin{subequations}
\begin{align} 
{\mathcal E}^\mu & =  R \, \gamma^\mu_\perp, & {\mathcal C}^\mu & =  {\mathcal E}^\mu (R \to \bar R), \\
 {\mathcal F}^{\mu \nu}  & =  R' \,  g_\perp^{\mu \nu}, &  {\mathcal G}^{\mu \nu}  & =  {\mathcal F}^{\mu \nu} (R' \to \bar R').
\end{align} 
\end{subequations}
Note the parameters $\bar R$ and $\bar R'$ appearing for the negative parity doublets. 
Calculating the matrix elements for all states given in eq.~\eqref{eq:heavymesonrep} (including the $\ell=2$ states)  yields for $\Gamma=\mathds{1}$ only the following the  non-vanishing matrix elements
\begin{subequations} \label{eq:appBiDQ}
\begin{align}
\Braket{B| \bar{b}_v   iD_\mu^\perp Q_v  | 1^+, \tfrac{1}{2} } &= -2 \sqrt{M_B M_E} R \, \eta_\mu , \\
\Braket{ B | \bar{b}_v  iD_\mu^\perp  Q_v  | 1^+, \tfrac{3}{2}} &= -2\sqrt{\frac{2}{3}} \sqrt{M_B M_{F}} \,R'  \, \eta_\mu,
\end{align}
\end{subequations}
while the non-vanishing matrix elements  containing $\Gamma= i\sigma^\perp_{\alpha\beta}$ are
\begin{subequations} 
\begin{align}
\Braket{B | \bar{b}_v   iD_\mu^\perp  i\sigma^\perp_{\alpha\beta}Q_v  | 1^-,\tfrac{1}{2} } &=  2 i \sqrt{M_B M_{C}} \, \bar R \,  (\epsilon_{\alpha\beta\mu\gamma}v^\gamma - 3 v_{[\alpha} \epsilon_{\beta\mu]\gamma}\eta^\gamma )  ,\\
\Braket{B | \bar{b}_v   iD_\mu^\perp  i\sigma^\perp_{\alpha\beta}Q_v  | 0^+,\tfrac{1}{2} } &=  2 i  \sqrt{M_B M_{E}} \, R \,  \epsilon_{\alpha\beta\mu\gamma}v^\gamma  ,\\
\Braket{B | \bar{b}_v   iD_\mu^\perp  i\sigma^\perp_{\alpha\beta}Q_v  | 1^+,\tfrac{1}{2} } &=  - 4  \sqrt{M_B M_{E}} \, R \,  \eta_{[\alpha}g^\perp_{\beta]\mu}  ,\\
\Braket{B | \bar{b}_v   iD_\mu^\perp  i\sigma^\perp_{\alpha\beta}Q_v  | 1^+,\tfrac{3}{2} } &=  2\sqrt{\frac{2}{3}}  \sqrt{M_B M_{F}} \, R' \,  \eta_{[\alpha}g^\perp_{\beta]\mu} , \\
\Braket{B | \bar{b}_v   iD_\mu^\perp  i\sigma^\perp_{\alpha\beta}Q_v  | 2^+,\tfrac{3}{2} } &=  i \sqrt{2}  \sqrt{M_B M_{F}} \, R' \,  \epsilon_{\alpha\beta\gamma\delta} \eta_{\mu}^{\,\,\,\gamma} v^\delta  ,\\
\Braket{B | \bar{b}_v   iD_\mu^\perp  i\sigma^\perp_{\alpha\beta}Q_v  | 1^-,\tfrac{3}{2} } &=  - i\sqrt{\frac{2}{3}}  \sqrt{M_B M_{G}} \, \bar R' \,  (\epsilon_{\alpha\beta\mu\gamma}v^\gamma - 3 v_{[\alpha} \epsilon_{\beta\mu]\gamma}\eta^\gamma ),
\end{align}
\end{subequations}
where we have denoted the polarization vectors and tensors by $\eta_\mu$ and $\eta_{\mu\nu}$, respectively, to avoid confusion with the Levi-Civita tensor $\epsilon_{\alpha\beta\mu\nu}$. Note, that because there are only two non-vanishing matrix elements for the $\Gamma=\mathds{1}$ case, only these two states will contribute in the expansion of the matrix elements at order $(\Lambda_{\rm QCD} / m_b)^2$ as stated in eq.~\eqref{eq:reducedsum}. Using the polarization sum rules from eq.~\eqref{eq:polprop}, then yields eq.~\eqref{eq:BiDiDGB}. Furthermore these matrix elements yield eq.~\eqref{eq:matrixmasterorder3},~\eqref{eq:matrixmasterorder4a} and~\eqref{eq:matrixmasterorder5a}.

\subsection{Dimension 5 matrix elements \label{app:detailsdim5}}
For the dimension five matrix elements we parametrize the light degrees of freedom by
\begin{subequations}
\begin{align} 
{\mathcal C}_{\mu\nu} &=  \frac{1}{3} \mu_\pi^2  g^\perp_{\mu\nu} - \frac{1}{6} \mu_G^2 i\sigma^\perp_{\mu\nu}, 
	& {\mathcal E}_{\mu\nu} & =  {\mathcal C}_{\mu\nu}(\mu_{\pi,g} \to \bar \mu_{\pi,g}), \\
{\mathcal G}_{\mu\nu\rho} &=  \lambda_S  g^\perp_{\rho\{\mu}\gamma^\perp_{\nu\}} + \lambda_A  g^\perp_{\rho[\mu}\gamma^\perp_{\nu]}, 
	& {\mathcal F}_{\mu\nu\rho} & = {\mathcal G}_{\mu\nu\rho} (\lambda_{A,S} \to \bar \lambda_{A,S}).
\end{align} 
\end{subequations}
where $\{\mu\nu\}$ denotes symmetrization in the indices $(\mu,\nu)$. Note again, that we that we have more parameters. But these  only appear in non-zero matrix elements which do not contribute in the LLSA approximation.
We obtain the non-vanishing spin-singlet matrix elements
\begin{subequations} \label{eq:5dimMEsinglet}
\begin{align} 
\Braket{B | \bar{b}_v   iD_\mu^\perp iD_\nu^\perp  Q_v  | 0^-, \tfrac{1}{2} } &= -\frac{2}{3} \sqrt{M_B M_{C}}\mu_\pi^2 g^\perp_{\mu\nu} , \\ 
\Braket{B | \bar{b}_v   iD_\mu^\perp iD_\nu^\perp  Q_v  | 1^-, \tfrac{1}{2} } &= \frac{1}{3} i \sqrt{M_B M_{C}}\mu_G^2 \epsilon^{\mu\nu\alpha\beta}v_\alpha\eta_\beta , \\
\Braket{B | \bar{b}_v   iD_\mu^\perp iD_\nu^\perp  Q_v  | 1^-, \tfrac{3}{2} } &= -i\sqrt{\frac{2}{3}} \lambda_A  \epsilon^{\mu\nu\alpha\beta}v_\alpha\eta_\beta , \\
\Braket{B | \bar{b}_v   iD_\mu^\perp iD_\nu^\perp  Q_v  | 2^-, \tfrac{3}{2} } &=- \sqrt{2} \lambda_S \eta_{\mu\nu} ,
\end{align}
\end{subequations}
and the non-vanishing spin-triplet matrix elements
\begin{subequations} \label{eq:5dimMEtriplet}
\begin{align}
\Braket{B | \bar{b}_v   iD_\mu^\perp iD_\nu^\perp i \sigma^\perp_{\alpha\beta} Q_v  | 0^-, \tfrac{1}{2} } &= -\frac{2}{3} \sqrt{M_B M_{C}} \mu_G^2  g^\perp_{\mu[\alpha}g^\perp_{\beta]\nu} , \\
\Braket{B | \bar{b}_v   iD_\mu^\perp iD_\nu^\perp i \sigma^\perp_{\alpha\beta} Q_v  | 1^-, \tfrac{1}{2} } & = \frac{2}{3}i\sqrt{M_B M_{C}} \\
\times
\Big\{   \mu_G^2 \big[ \eta_\mu\epsilon_{\alpha\beta\nu\delta} & -\eta_{[\alpha}\epsilon_{\beta]\mu\nu\delta}  - 2 g^\perp_{\mu[\alpha}\epsilon_{\beta]\nu\gamma\delta}\eta^\gamma  \big]_{\mu\nu} + \mu_\pi^2 g^\perp_{\mu\nu} \epsilon_{\alpha\beta\gamma\delta} \eta^\gamma \Big\} v^\delta \nonumber  , \\
\Braket{B | \bar{b}_v   iD_\mu^\perp iD_\nu^\perp i \sigma^\perp_{\alpha\beta} Q_v  | 0^+, \tfrac{1}{2} } &= -\frac{1}{3} i \sqrt{M_B M_{E}} \bar \mu_G^2 ( \epsilon_{\mu\nu\alpha\beta} + 2 v_{[\mu}\epsilon_{\nu]\alpha\beta\gamma}v^\gamma + 2 v_{[\alpha}\epsilon_{\beta]\mu\nu\gamma}v^\gamma )  , \\
\Braket{B | \bar{b}_v   iD_\mu^\perp iD_\nu^\perp i \sigma^\perp_{\alpha\beta} Q_v  | 2^+, \tfrac{3}{2} } &= \frac{1}{3} i \sqrt{M_B M_{F}} 
	\Big\{ \bar \lambda_A ( \epsilon_{\alpha\beta\gamma[\mu} + 2 v^\delta v_{[\alpha}\epsilon_{\beta]\gamma\delta[\mu} - \epsilon_{\alpha\beta\gamma\delta}v^\delta v_{[\mu} ) \eta_{\nu]}^{\,\,\,\gamma} \nonumber \\
	& \qquad \qquad    -	\bar \lambda_S ( \epsilon_{\alpha\beta\gamma\{\mu} + 2 v^\delta v_{[\alpha}\epsilon_{\beta]\gamma\delta\{\mu} - \epsilon_{\alpha\beta\gamma\delta}v^\delta v_{\{\mu} ) \eta_{\nu\}}^{\,\,\,\gamma} \Big\}  , \\
\Braket{B | \bar{b}_v   iD_\mu^\perp iD_\nu^\perp i \sigma^\perp_{\alpha\beta} Q_v  | 1^-, \tfrac{3}{2} } & = i \sqrt{\frac{2}{3}}\sqrt{M_B M_{G}} \\
\times
\Big\{   \lambda_A \big[ \eta_\mu\epsilon_{\alpha\beta\nu\delta} & + 2\eta_{[\alpha}\epsilon_{\beta\mu\nu\delta}  + 4  g^\perp_{\mu[\alpha}\epsilon_{\beta]\nu\gamma\delta}\eta^\gamma  \big]_{\mu\nu}  +\lambda_S \big( 3 \eta_{\{\mu}  \epsilon_{\nu\}\alpha\beta\delta} - g^\perp_{\mu\nu}\epsilon_{\alpha\beta\gamma\delta} \eta^\gamma  \Big\} v^\delta \nonumber  , \\
\Braket{B | \bar{b}_v   iD_\mu^\perp iD_\nu^\perp i \sigma^\perp_{\alpha\beta} Q_v  | 2^-, \tfrac{3}{2} } & =  2 \sqrt{2} \sqrt{M_B M_{G}}   \big[ -  \lambda_A g^\perp_{\alpha[\mu}\eta_{\nu]\beta}  + \lambda_S g^\perp_{\alpha\{\mu}\eta_{\nu\}\beta}  \big]_{\alpha\beta}  ,
\end{align}
\end{subequations}
where   $[...]_{\alpha\beta}$ denote antisymmetrization like $[\alpha\beta]$, i.e. $\left[ T_{\alpha\beta}\right] = T_{[\alpha\beta]}$. 
The matrix elements for the $j=3/2$ states do not contribute in LLSA and
the remaining matrix elements  are used to obtain eq.~\eqref{eq:BiDiDiDiDB} and eq.~\eqref{eq:matrixmasterorder5b}.

\subsection{Dimension 6 matrix elements containing $(iv\cdot D)$ \label{app:detailsdim6}}
The light brown muck is  parametrized by
\begin{subequations}
\begin{align} 
{\mathcal C}_{\mu\nu} &=   -\frac{1}{3} \rho_D^3  g^\perp_{\mu\nu} - \frac{1}{6} \rho_{LS}^3 i\sigma^\perp_{\mu\nu},
	& {\mathcal E}_{\mu\nu} & =  {\mathcal C}_{\mu\nu}(\rho_{D,LS} \to \bar \rho_{D,LS}), \\
{\mathcal G}_{\mu\nu\rho} &=  \kappa_S  g^\perp_{\rho\{\mu}\gamma^\perp_{\nu\}} + \kappa_A  g^\perp_{\rho[\mu}\gamma^\perp_{\nu]},
	& {\mathcal F}_{\mu\nu\rho} & = {\mathcal G}_{\mu\nu\rho} (\kappa_{A,S} \to \bar \kappa_{A,S}).
\end{align} 
\end{subequations}
The non-vanishing matrix elements are completely analogous to the five-dimensional ones given in eq.~\eqref{eq:5dimMEsinglet} and~\eqref{eq:5dimMEtriplet},
\begin{subequations}
\begin{align}
\Braket{B | \bar{b}_v   iD_\mu^\perp (iv\cdot D)iD_\nu^\perp  Q_v  | 0^-, \tfrac{1}{2} } &= \frac{2}{3} \sqrt{M_B M_{C}}\rho_D^3 g^\perp_{\mu\nu} , \\ 
\Braket{B | \bar{b}_v   iD_\mu^\perp(iv\cdot D) iD_\nu^\perp  Q_v  | 1^-, \tfrac{1}{2} } &= \frac{1}{3} i \sqrt{M_B M_{C}}\rho_{LS}^3 \epsilon^{\mu\nu\alpha\beta}v_\alpha\epsilon_\beta , \\
\Braket{B | \bar{b}_v   (iD_\mu^\perp)(iv\cdot D) (iD_\nu^\perp)  Q_v  | 1^-, \tfrac{3}{2} } &= -i\sqrt{\frac{2}{3}} \kappa_A  \epsilon^{\mu\nu\alpha\beta}v_\alpha\epsilon_\beta ,  \\
\Braket{B | \bar{b}_v   (iD_\mu^\perp)(iv\cdot D) (iD_\nu^\perp)  Q_v  | 2^-, \tfrac{3}{2} } &=- \sqrt{2} \kappa_S \epsilon_{\mu\nu} ,
\end{align}
\end{subequations}
and
\begin{subequations}
\begin{align}
\Braket{B | \bar{b}_v   iD_\mu^\perp(iv\cdot D) iD_\nu^\perp i \sigma^\perp_{\alpha\beta} Q_v  | 0^-, \tfrac{1}{2} } &= -\frac{2}{3} \sqrt{M_B M_{C}} \rho_{LS}^3  g^\perp_{\mu[\alpha}g^\perp_{\beta]\nu} , \\
\Braket{B | \bar{b}_v   iD_\mu^\perp(iv\cdot D) iD_\nu^\perp i \sigma^\perp_{\alpha\beta} Q_v  | 1^-, \tfrac{1}{2} } & = \frac{2}{3}i\sqrt{M_B M_{C}}   \\
\times
\Big\{   \rho_{LS}^3 \big[ \epsilon_\mu\epsilon_{\alpha\beta\nu\delta} & -\epsilon_{[\alpha}\epsilon_{\beta]\mu\nu\delta}  - 2 g^\perp_{\mu[\alpha}\epsilon_{\beta]\nu\gamma\delta}\epsilon^\gamma  \big]_{\mu\nu} - \rho_D^3 g^\perp_{\mu\nu} \epsilon_{\alpha\beta\gamma\delta} \epsilon^\gamma \Big\} v^\delta \nonumber  , \\
\Braket{B | \bar{b}_v   iD_\mu^\perp(iv\cdot D)  iD_\nu^\perp i \sigma^\perp_{\alpha\beta} Q_v  | 0^+, \tfrac{1}{2} } &= -\frac{1}{3} i \sqrt{M_B M_{E}} \bar \rho_{LS}^3 ( \epsilon_{\mu\nu\alpha\beta} + 2 v_{[\mu}\epsilon_{\nu]\alpha\beta} + 2 v_{[\alpha}\epsilon_{\beta]\mu\nu} )  , \\
\Braket{B | \bar{b}_v   iD_\mu^\perp (iv\cdot D) iD_\nu^\perp i \sigma^\perp_{\alpha\beta} Q_v  | 2^+, \tfrac{3}{2} } &= \frac{1}{3} i \sqrt{M_B M_{F}} 
	\Big\{ \bar \kappa_A ( \epsilon_{\alpha\beta\gamma[\mu} + 2 v^\delta v_{[\alpha}\epsilon_{\beta]\gamma\delta[\mu} - \epsilon_{\alpha\beta\gamma\delta}v^\delta v_{[\mu} ) \eta_{\nu]}^{\,\,\,\gamma} \nonumber \\
	& \quad    -\bar \kappa_S ( \epsilon_{\alpha\beta\gamma\{\mu} + 2 v^\delta v_{[\alpha}\epsilon_{\beta]\gamma\delta\{\mu} - \epsilon_{\alpha\beta\gamma\delta}v^\delta v_{\{\mu} ) \eta_{\nu\}}^{\,\,\,\gamma} \Big\}    , \\
\Braket{B | \bar{b}_v   iD_\mu^\perp (iv\cdot D)iD_\nu^\perp i \sigma^\perp_{\alpha\beta} Q_v  | 1^-, \tfrac{3}{2} } & = i \sqrt{\frac{2}{3}}\sqrt{M_B M_{G}} \\
\times
\Big\{   \kappa_A \big[ \epsilon_\mu\epsilon_{\alpha\beta\nu\delta} + 2\epsilon_{[\alpha}\epsilon_{\beta\mu\nu\delta}  & + 4  g^\perp_{\mu[\alpha}\epsilon_{\beta]\nu\gamma\delta}\epsilon^\gamma  \big]_{\mu\nu}  +\kappa_S \big( 3 \epsilon_{\{\mu}  \epsilon_{\nu\}\alpha\beta\delta} - g^\perp_{\mu\nu}\epsilon_{\alpha\beta\gamma\delta} \epsilon^\gamma )  \Big\} v^\delta \nonumber  , \\
\Braket{B | \bar{b}_v   iD_\mu^\perp(iv\cdot D) iD_\nu^\perp i \sigma^\perp_{\alpha\beta} Q_v  | 2^-, \tfrac{3}{2} } & =  2 \sqrt{2} \sqrt{M_B M_{G}}   \big[ -  \kappa g^\perp_{\alpha[\mu}\epsilon_{\nu]\beta}  + \kappa_S g^\perp_{\alpha\{\mu}\epsilon_{\nu\}\beta}  \big]_{\alpha\beta}  .
\end{align}
\end{subequations}
These matrix elements then contribute to eq.~\eqref{eq:matrixmasterorder5b}.

\end{appendix}

\providecommand{\href}[2]{#2}\begingroup\raggedright\endgroup



%

\end{document}